\documentclass[conference]{content}
\IEEEoverridecommandlockouts
\usepackage{cite}

\usepackage{amsmath,amssymb,amsfonts,mathtools,bm}
\usepackage{amsthm}
\usepackage{algorithm}
\usepackage{algorithmic}
\usepackage{graphicx}
\usepackage{textcomp}
\usepackage{xcolor,float}
\usepackage{tabularx}
\usepackage{textcomp}
\usepackage{diagbox}
\usepackage{svg}
\usepackage{booktabs}
\usepackage{multirow}
\usepackage{makecell}
\usepackage{bbding}
\usepackage{hyperref}

\allowdisplaybreaks
\renewcommand{\baselinestretch}{1}

\begin{document}
    
    \title{{SigT: An Efficient End-to-End MIMO-OFDM Receiver Framework Based on Transformer}}
\author{
\IEEEauthorblockN{
Ziyou Ren\IEEEauthorrefmark{1},
Nan Cheng\IEEEauthorrefmark{1},
Ruijin Sun\IEEEauthorrefmark{1}, 
Xiucheng Wang\IEEEauthorrefmark{1},
Ning Lu\IEEEauthorrefmark{2} and
Wenchao Xu\IEEEauthorrefmark{3}\\
}
\IEEEauthorblockA{
\IEEEauthorrefmark{1}School of Telecommunications Engineering,
Xidian University, Xi'an, China\\
\IEEEauthorrefmark{2}Department of Electrical and Computer Engineering, Queen’s University, Kingston K7L 3N6, Ontario, Canada\\
\IEEEauthorrefmark{3}Department of Computing, The Hong Kong Polytechnic University, Hongkong, China\\
Email: \{zyren\_681,xcwang\_1\}@stu.xidian.edu.cn, dr.nan.cheng@ieee.org,sunruijin@xidian.edu.cn\\  ning.lu@queensu.ca, wenchao.xu@polyu.edu.hk
}}

    \maketitle

\IEEEdisplaynontitleabstractindextext

\IEEEpeerreviewmaketitle

\begin{abstract}
Multiple-input multiple-output and orthogonal frequency-division multiplexing (MIMO-OFDM) are the key technologies in 4G and subsequent wireless communication systems. Conventionally, the MIMO-OFDM receiver is performed by multiple cascaded blocks with different functions and the algorithm in each block is designed based on ideal assumptions of wireless channel distributions. However, these assumptions may fail in practical complex wireless environments. The deep learning (DL) method has the ability to capture key features from complex and huge data. In this paper, a novel end-to-end MIMO-OFDM receiver framework based on \textit{transformer}, named SigT, is proposed. By regarding the signal received from each antenna as a token of the transformer, the spatial correlation of different antennas can be learned and the critical zero-shot problem can be mitigated. Furthermore, the proposed SigT framework can work well without the inserted pilots, which improves the useful data transmission efficiency. Experiment results show that SigT achieves much higher performance in terms of signal recovery accuracy than benchmark methods, even in a low SNR environment or with a small number of training samples. Code is available at \url{https://github.com/SigTransformer/SigT}. 

\end{abstract}

\begin{IEEEkeywords}
Deep learning, Transformer, MIMO-OFDM, self-attention.

\end{IEEEkeywords}

\section{Introduction}
The multiple-input multiple-output and orthogonal frequency-division multiplexing (MIMO-OFDM) are key technologies in the physical layer for 4G and 5G broadband wireless systems and will keep playing an important role in 6G wireless systems. Benefiting from the beamforming gain, the diversity gain, and the multiplexing gain, MIMO technology can significantly increase the system spectral efficiency and reliability \cite{larsson2014massive}. By transforming the digital signal from the time domain to the frequency domain to transmit, OFDM technology can effectively deal with the frequency-selective property of the broadband channel \cite{stuber2004broadband}. With the MIMO-OFDM technology, the conventional receiver consists of multiple cascaded  blocks, i.e., fast Fourier transformation (FFT), MIMO channel estimation (CE), MIMO signal detection (SD), and demodulation. These blocks are designed independently to perform different functions and algorithms, which are based on ideal assumptions of channel distributions and linear interference. However, in practice, these assumptions may fail due to the complex wireless environment and the non-linear effect of hardware circuits, resulting in deteriorated bit error rate (BER) performance. 
Deep learning (DL) techniques \cite{NIPS2012_c399862d}, which has achieved great success in the field of natural language processing (NLP) and computer vision (CV), have been widely applied to the field of communication in recent years \cite{9882279, wang2022joint}. Since DL can capture the complex and non-linear features of wireless channels by feeding with huge labeled data, it can be a potential way to solve these issues. To reap this advantage, many researchers introduced the DL technologies into the physical layer of communications and studied the DL-based OFDM receiver \cite{10,11,13,12}. Both DL-based individual receiver block, such as CE \cite{8715649}, MIMO SD \cite{10}, channel state information feedback \cite{13}, and autoencoder-based end-to-end communication system \cite{8214233}, and DL-based end-to-end OFDM receiver \cite{gao2018comnet, 8715649, 12} have been investigated. For the DL-based individual receiver blocks, the signal recovery process can be split into subnets. Each subnet is designated to serve the function of a conventional receiver's block, e.g., CE and SD, between which hand-crafted information are inserted and locally optimizes the subnet. 
However, this framework is based on expert knowledge and requires sophisticated design. Therefore, the DL-based end-to-end receiver framework is recently proposed and studied where little expert knowledge is exploited.

For the DL-based end-to-end receiver, the signal recovery process is regarded as a whole block. It can be optimized from a global perspective, which can deal with the complicated wireless environment and the mismatch of block splitting in the conventional receiver. 
In \cite{11}, an end-to-end DL-based receiver, named FC-DNN, is explored, which maps the received OFDM subcarriers to constellation symbols. Although such a network performs well in signal detection tasks, expert knowledge is still required for quadrature amplitude modulation (QAM) demodulation block. FC-DNN is further evaluated in \cite{12}, which reveals that FC-DNN performs well even in mismatched channels and achieves competitive performance compared to DL-based individual-block cascading networks. However, compared with the DL-based individual-block cascading receiver design, the DL-based end-to-end method treats the signal recovery in the receiver as a black box, and cannot exploit the abundant expert knowledge in the field of communication systems. This leads to the fact that when the length of input digits increases, the possibility of recovered signal increases exponentially, and so does the signal recovery difficulty. This issue is often referred to as zero-shot problem \cite{NIPS2013_2d6cc4b2}. Especially when it comes to the MIMO case, the issue gets worse since the input digit sequence is usually much longer than in the single-input single-output (SISO) case. 

Furthermore, existing literature reveals that the structure of the neural network significantly affects the convergence and performance of the DL-based models. Except for the FC-DNN and convolutional neural network (CNN) adopted in the literature, many emerging DL structures have shown powerful ability in NLP and CV. In \cite{vaswani2017attention}, a novel sequence-to-sequence (Seq2Seq) neural model for machine translation, \textit{transformer}, is proposed and achieves state-of-the-art performance. In \cite{dosovitskiy2020image}, Vision Transformer (ViT) is proposed to extend transformer to computer vision tasks, e.g., object recognition, by employing transformer encoder as the model backbone. 

Motivated by the discussion above, in this paper, we propose a Transformer-based DL model to design an end-to-end MIMO-OFDM receiver, named signal Transformer (SigT). By treating the signal received at each antenna as a token of the Transformer framework, the spatial correlation of different antennas can be learned and the critical zero-shot problem can be mitigated. The main contributions of this paper are summarized as follows.
\begin{itemize}
    \item A novel neural network framework called SigT is proposed in this paper to design the end-to-end MIMO-OFDM receiver. 
    The proposed SigT can serve as a reference and baseline for future study of end-to-end learning-based MIMO-OFDM receiver design. 
    \item Compared with most existing DL-based receivers, our proposed SigT MIMO-OFDM receiver can recover the data signal with high accuracy without any foreknown pilots. By saving the time-frequency resource occupied by pilots, the proposed SigT-based wireless communication systems can transmit more useful data signals in each frame and thus enlarge the effective channel capacity. 
    \item  We thoroughly evaluate the performance of the proposed SigT through extensive experiments with a real dataset. Experiment results show that SigT outperforms existing DL-based end-to-end models by a large margin.
\end{itemize}

The remainder of the paper is organized as follows. Section II gives the preliminaries on the MIMO-OFDM system and describes the task of end-to-end MIMO-OFDM receiver design. In Section III, the proposed SigT framework is described in detail. In Section IV, the performance of SigT is evaluated through experiments. Finally, Section V concludes the paper. 

\begin{figure*}
  \centering
  \includegraphics[width=1.75\columnwidth]{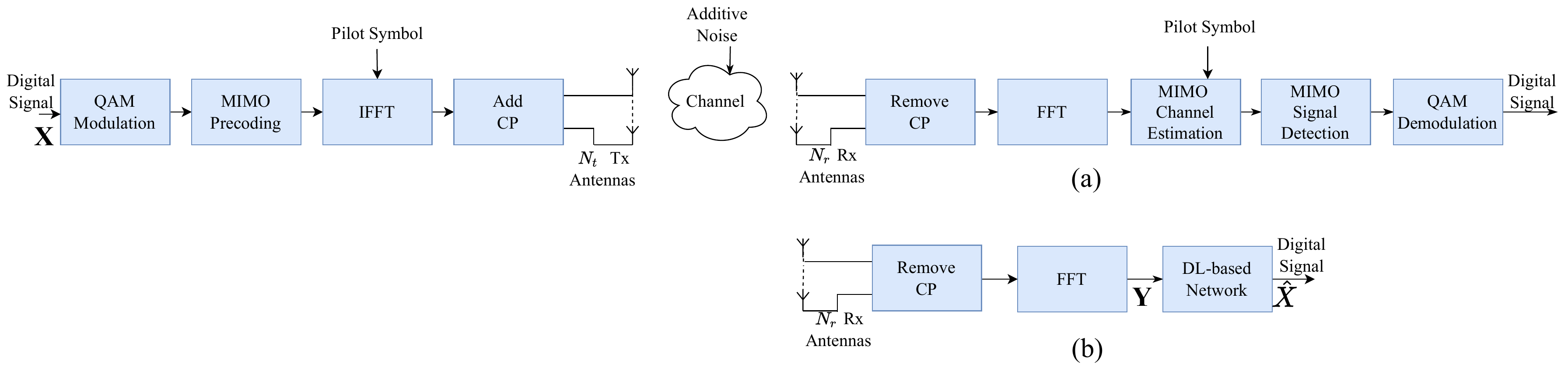}
   \vspace{-10pt}
  \caption {Block diagram of MIMO-OFDM system. (a) is the conventional MIMO-OFDM receiver system (b) is the receiver design of an end-to-end DL-based model.}
  \label{system}
   \vspace{-20pt}
\end{figure*}

\section{Preliminaries and Problem Statement}
In this section, we first shortly introduce the preliminaries on the MIMO-OFDM system and DL-based receiver. Then, we describe the specific task of end-to-end DL-based MIMO-OFDM receiver design.

\subsection{MIMO-OFDM System}
As illustrated in Fig. \ref{system}(a), a typical MIMO-OFDM system includes $N_t$-antenna transmitter, $N_r$-antenna receiver, $N_s$ subcarriers , and $N_i$ information in each subcarrier.
For the baseband process in the transmitter, the digital bit stream $X$ successively goes through the QAM modulation block, the MIMO precoding block, the IFFT block, and the CP adding a block. During the QAM modulation, the bit stream is modulated as the  data symbol stream in complex form. Then, the MIMO precoding is designed to map the data symbol stream to the transmitter antennas. In each antenna, the useful data symbol sequence and a few pilot symbols are grouped together to perform the IFFT operation and generate the OFDM symbol. Thereafter, the CP is added before each OFDM symbol to combat the inter-symbol interference (ISI) caused by the multi-path effect of wireless channels.

In the conventional MIMO-OFDM receiver, the received time-domain signal goes through the following blocks in turn: the CP removing, the FFT, the MIMO CE, the MIMO signal detection, and the QAM demodulation. In particular, in each receiver antenna, the CP contaminated by the ISI is first removed  and the FFT is performed to transform the received time-domain signal to the frequency domain. Then, the received pilot symbols of $N_r$ antennas are used to estimate the wireless MIMO channel with the dimension of $N_r \times N_t$. With the estimated MIMO channel, the received data symbols of $N_r$ antennas are detected via the receiver precoding. Thereafter, the detected symbol stream is sent to the QAM demodulation block to recover the original binary bit stream $\tilde X$.




\begin{figure}
  \centering
  \includegraphics[width=0.75\columnwidth]{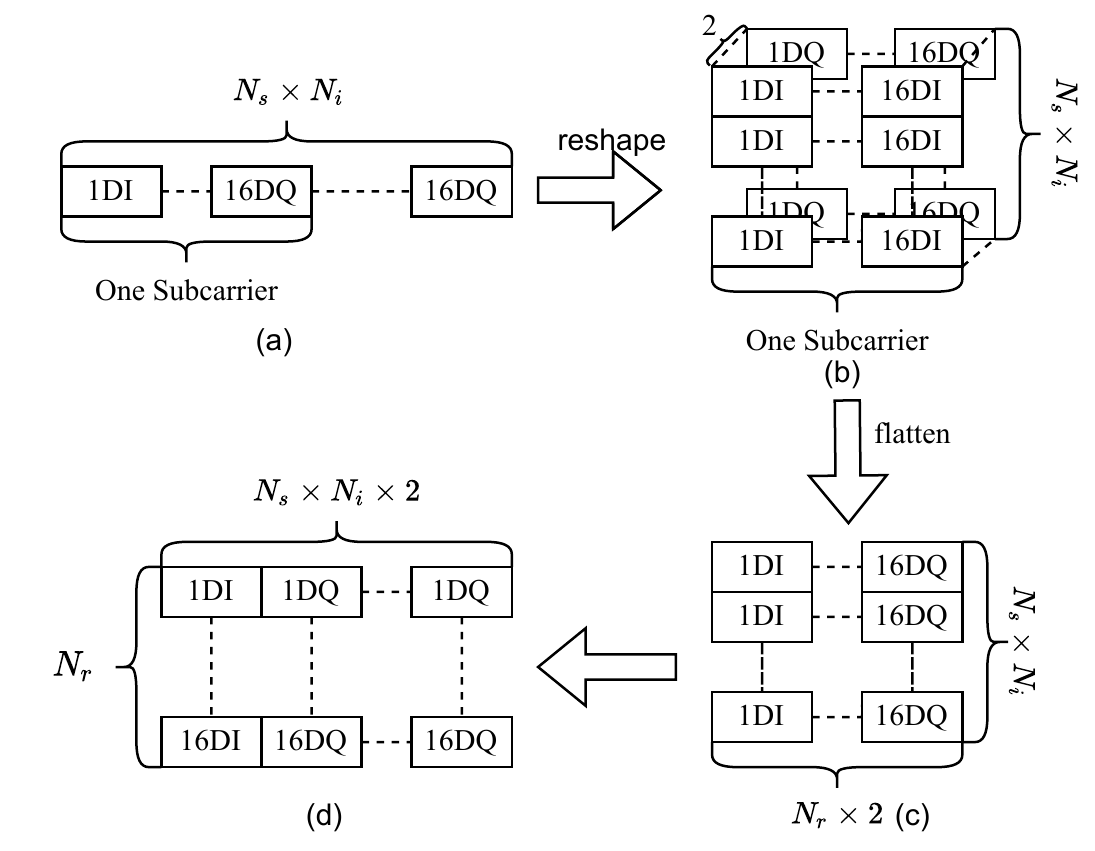}
  \vspace{-11pt}
  \caption {(a) Visualization of the input signal of AI network, i.e. $Y$ in our denotation. We plot using $N_r=16,\ N_t=4$ for demonstration. (b) we reshape the signal by corresponding each row to one of the 256 subcarriers. (c) we flatten the $3D$ tensor to be $2D$ to meet the input requirement of the Transformer module. (d) transpose the tensor and split each row as a token, which contains the information of an antenna, to feed into our network. }
  \label{data}
  \vspace{-11pt}
\end{figure}

\subsection{End-to-End MIMO-OFDM Receiver Design}

In this part, we describe the task of end-to-end DL-based MIMO-OFDM receiver design, as shown in Fig. \ref{system}(b). The DL-based model serves as a receiver block that substitutes CE, SD, and QAM demodulation in the conventional MIMO-OFDM system. The input of the model $Y\in \mathbb{R}^{N_s\times N_r\times N_i\times 2}$ after FFT is fed into DL-based model and the model directly outputs the predicted label $\hat{X}\in  \mathbb{R}^{N_s\times N_t\times 2}$. $\hat{X}$ is a string of float numbers predicted during the training phase and is used to compute mean square error (MSE) loss during the model training phase. In the test phase, the model outputs $\Tilde{X}$, which is a bitstream after applying a hard decision to $\hat{X}$. The dimension of $\hat{X}$ and $\Tilde{X}$ should be the same as $X$, i.e., $\hat{X}, \Tilde{X}\in \mathbb{R}^{N_s\times N_t\times 2}$.

For the performance of the MIMO-OFDM receiver, we follow the average accuracy (AACC) criteria. AACC is calculated using the following equation:
\begin{align*}
    & AACC = 1-\sum\limits_{n=1}^N\frac{|\Tilde{X}_n-X_n|}{N}, \tag{1}
\end{align*}
where $N$ is the total number of test signals, $\Tilde{X_n}$ and $X_n$ is the $n_{th}$ signal in the test set.

In end-to-end receiver design, the model needs to complete tasks of CE, SD, and QAM demodulation. This is far more complicated and difficult than simply fitting the function of an individual module. Since no expert knowledge is exploited, the model learns by making the predicted label approaching the ground truth. However, in this task, the dimension of the ground truth labels increases exponentially with the bitstream length, i.e., $2^{N_s\times N_t\times 2}$. Therefore, it is extremely hard for the dataset to cover all labels due to both the hardware storage limitation and the cost to collect the data, which requires the model to preserve the capability of predicting labels that are not included in the training set, leading to the zero-shot problem. The zero-shot problem may result in a large gap between training accuracy and test accuracy, which is referred to as overfitting. In the sequel, we design a transformer-based model, named SigT, to better learn the correlation among antennas and other higher semantics to improve the model generalization performance. 

\section{Signal Transformer}
\begin{figure*}
  \centering
  \includegraphics[width=1.7\columnwidth,height=5.5cm]{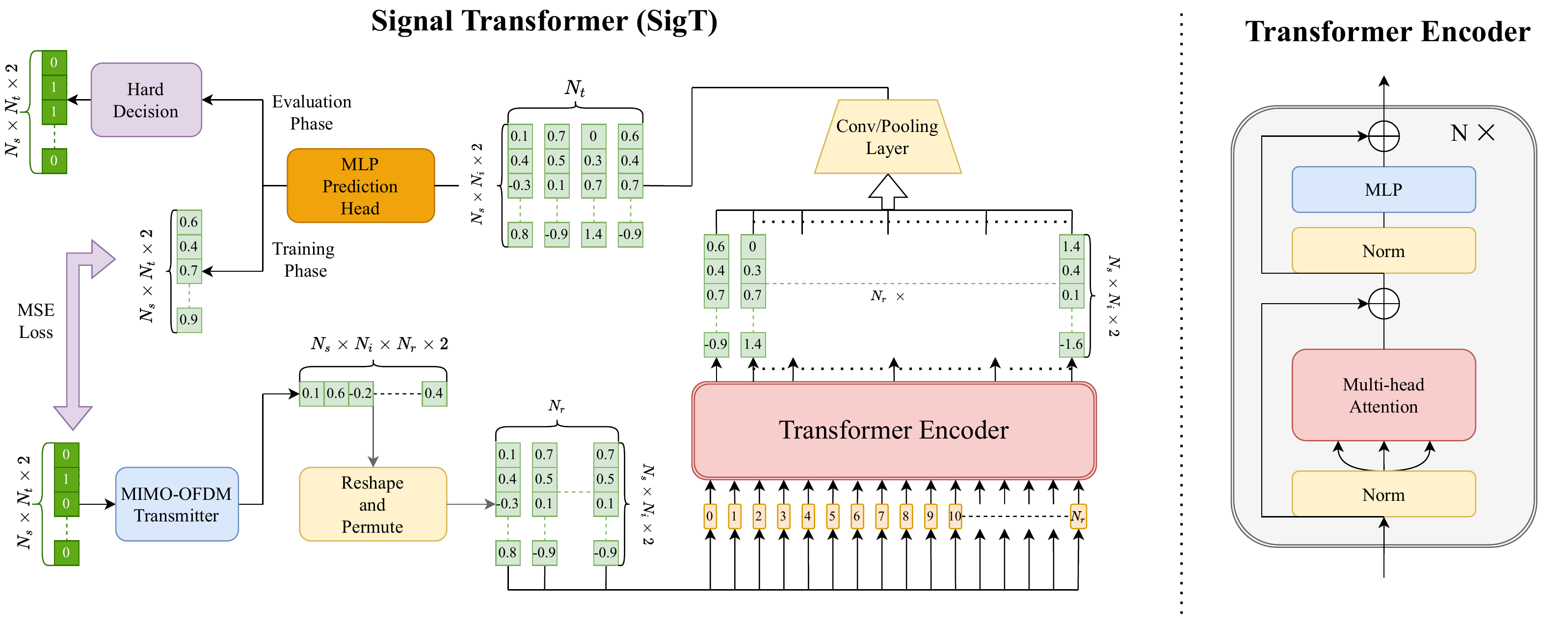}
   \vspace{-17pt}
  \caption {Model review (left). We first reshape and permute the input signal, and retain all subcarriers from one antenna to a single vector. Then the processed signal is fed into the transformer Encoder, which will output a sequence of features each has the same dimension as the inputs. After that, we reduce the number of vectors by going through pooling layers or convolutional layers to $N_t$ vectors. Finally, we concatenate $N_t$ vectors and pass through an MLP prediction head to extract the features and get the target vector that has the same dimension as the ground truth signal. Note that during the test phase, we apply an additional hard decision layer to convert predicted float numbers to a bitstream. The detail of the Transformer Encoder (right) is inspired by \cite{vaswani2017attention}.}
  \label{SigT}
   \vspace{-11pt}
\end{figure*}

In this section, we first briefly introduce the multi-head self-attention mechanism (MSA), which is the key module of the transformer. Then, we propose a transformer-based end-to-end MIMO-OFDM receiver model, SigT, and give the design details. 


\subsection{Multi-Head Self-Attention Mechanism}

Self-attention is a basic block in Transformer, which captures the feature for each vector and aggregates them together to get an output set of features that has the same dimension and number as the input set of vectors. The pivotal issue is to get query ($Q$), key ($K$), and value ($V$) for each input vector by applying a linear transformation. Then, we can compute the attention by the following equations.
\begin{align*}
    & Q = W^Q\cdot I, W^Q\in \mathbb{R}^{d_Q\times d}, \tag{2a} \\
    & K = W^K\cdot I, W^K\in \mathbb{R}^{d_K\times d}, \tag{2b} \\
    & V = W^V\cdot I, W^V\in \mathbb{R}^{d_V\times d}, \tag{2c} \\
    & \text{Attention}(I) = V\cdot \text{Softmax}(\frac{K^TQ}{\sqrt{d_K}}), \tag{2d}
\end{align*}
where $I$ is the tensor concatenated by the input sequence of vectors, $d$ is the dimension of input vectors, $d_Q,d_K,d_V$ are dimension of $Q,K,V$, respectively. Normally, we set $d_Q=d_K$ such that we can scaled $K^TQ$ by multiplying it with $\frac{1}{\sqrt{d_K}}$.

Then, a multi-head attention block can be built based on the aforementioned self-attention block.
\begin{align*}
    & \text{MultiHead} = W^O\cdot \text{Concat}(head_1,head_2,\cdots,head_h), \tag{3a} \\
    & head_i = \text{Attention}(I_i), \tag{3b} 
\end{align*}

\subsection{Signal Transformer Model}
In MIMO systems, each receiver antenna receives the signal from all transmitting antennas. Thereby, the received signals of each receiver antenna are strongly correlated, and this correlation plays an important role in recovering the signal.
To capture this correlation, we propose SigT, which is based on the transformer, where the correlation among antennas is learned by self-attention during signal recovery. The overall architecture of SigT is shown in Fig. \ref{SigT}.

\subsubsection{Tokenization}
The standard Transformer model takes in a sequence of token embeddings, then outputs features for each token. 
To better fit the transformer model and learn the correlation among antennas, proper tokenization is required. In Signal transformer, the input data is permuted and reshaped to form the tokens, as shown in Fig. \ref{data}.
Given a high dimension vector, we first need to reshape it according to its physical properties, e.g. number of antennas, number of subcarriers, etc., then we reduce the dimension to form a $2D$ tensor. Finally, we split the tensor on the dimension of $N_r$ into $N_r$ vectors, each serves as a token, and feed the sequence of tokens into the transformer Encoder module.

\subsubsection{Transformer Encoder}
After tokenization, token sequences are fed into the transformer Encoder. The design of the transformer Encoder is the same as the original design \cite{vaswani2017attention}. This structure is able to capture the relation between different tokens, which, in this circumstance, complements missing information of some antennas and enhances the important information shared among antennas. 

Normally, MSA performs better than single-head attention while not increasing asymptotic complexity. This is because multiple types of relations can be learned rather than only one type of relation in the single-head attention mechanism.

    

\subsubsection{Aggregation}
The output features of the transformer Encoder are passed through pooling layers or convolutional layers to aggregate the information from nearby features. Here, we denote the output features of transformer Encoder, i.e., the input features of pooling/convolutional layers, by $f_i\in \mathbb{R}^{N_s\times N_i\times N_r\times 2},i=1,2,\cdots,N_r$, and the output features of pooling/convolutional layer by $p_i\in \mathbb{R}^{N_s\times N_t\times 2},i= 1,2,\cdots,N_t$, respectively.

\subsubsection{Output and Loss Calculation}
Finally, we concatenate $p_i\in \mathbb{R}^{N_s\times N_t\times 2},i= 1,2,\cdots,N_t$ to form $P$ and feed it into a two-layer MLP to get $\hat{X}$ as follows.
\begin{align*}
    & P = \text{Concat}(p_1,p_2,\cdots,p_{N_t}),\quad P\in \mathbb{R}^{N_s\times N_t\times 2}, \tag{4a}\\
    & \hat{X} = \text{Sigmoid}(\text{MLP}(P)),\quad \hat{X}\in \mathbb{R}^{N_s\times N_t\times 2}, \tag{4b}\\
    & \Tilde{X} = \text{Hardmax}(\hat{X}), \quad \Tilde{X}\in \mathbb{R}^{N_s\times N_t\times 2}, \tag{4c}
\end{align*}
where $P$ denotes the input features, and $\hat{X}$ denotes the output features activated by the $Sigmoid$ function, which maps the real number to the range from $0$ to $1$. $\Tilde{X}$ is a bitstream which is the hard decision of $\hat{X}$.

During the training phase, we update the weights in MLP, conv/pooling layer, and transformer Encoder by computing mean square error (MSE) loss between $\hat{X}$ and $Y$ as $Loss = \frac{1}{N_g}\sum\limits_{i=1}^{N_g}(\hat{x_i}-x_i)^2$, where $N_g = 2 N_s N_t$ denotes the number of digits of signals, and $x_i$ and $\hat{x_i}$ denote the $i_{th}$ bit in $X$ and $\hat{X}$, respectively.

\section{Experiment and Discussion}
In this section, we first describe the dataset, experiment configurations, and two DL-based models which are used as benchmarks. Then, experiment results are shown and discussed to evaluate the performance of the proposed SigT.

\subsection{Experiment Configurations}
\begin{table*}[h]
	\caption{Model Structure Exploration}
	\label{table}
	\centering
	\setlength{\tabcolsep}{5pt}
        \vspace{-8pt}
	\renewcommand\arraystretch{0.95} 
	\begin{tabular}{|m{3cm}<{\centering}|m{1.2cm}<{\centering}|m{1.2cm}<{\centering}|m{1.2cm}<{\centering}|m{1.5cm}<{\centering}|m{1.2cm}<{\centering}|m{1.2cm}<{\centering}|m{1.2cm}<{\centering}|}
		\hline 
		Model backbone& Conv Layer & Pooling Layer & Dropout & Optimizer & Train SNR (dB)& \textbf{Train AACC (\%)} & \textbf{test aacc (\%)} \\ \hline
		\multirow{6}*{Transformer Encoder} &  & \checkmark &  & Adam & 10& 96.87& 76.07\\ \cline{2-8}
		 & \checkmark & & & Adam & 10 & \textbf{98.31} & \textbf{79.54}\\ \cline{2-8}
		 & & \checkmark &  & SGD & 10 &72.89 &70.27\\ \cline{2-8}
		 & &\checkmark & \checkmark&Adam &10 & 85.57&69.42 \\ \cline{2-8}
		 & \checkmark& &\checkmark &Adam &10 &89.48 &71.79\\ \cline{2-8}
		 & \checkmark& & &Adam &25 &None &79.28\\ \cline{2-8}
		 &  & \checkmark &  & Adam & 25& None & 76.23\\ \hline
		\multirow{2}*{LSTM} & \checkmark &  &  & Adam & 10& 58.5& 50.61\\ \cline{2-8}
		& \checkmark &  & \checkmark & Adam & 10& 57.75 & 50.86\\ \hline
	\end{tabular}
	\label{tab1e}
        \vspace{-15pt}
\end{table*}
To evaluate the performance of the proposed SigT, we use the channel data provided by Peng Cheng Laboratory (PCL) in national artificial intelligence competition (NAIC)\footnote{https://naic.pcl.ac.cn/contest/10/34}. The channel data is obtained in typical city area. The central frequency of the system is $3.5$ GHz, and was sampled from $500$ locations, each with $20$ measurements. We use this open data to generate our dataset.
\subsection{Performance Comparison with Benchmark Methods}

In the experiment, we set the number of subcarriers for OFDM to $N_s = 256$, number of transmitting antennas to $N_t=4$, number of receiving antennas to $N_r=16$. In each subcarrier, we have one information symbol for each antenna, that is $N_i = 1$. An illustration of serial signal $Y$ is depicted in Fig. \ref{data}, where we simply permute and reshape the original data to fit our model and without any additional transformations.

To construct our dataset, we keep the size of test set over the size of training set to be $1/10$. By default, the training set has $25600$ signals, and the test set has $2560$ signals. To load data more efficiently, we set the size of minibatch to be $640$, i.e., loading $640$ signals per iteration, and control the size of dataset using the number of minibatches, which is denoted by NB in Fig. \ref{EXP_3}.
\begin{figure}[h]
  \centering
  \includegraphics[width=0.75\columnwidth]{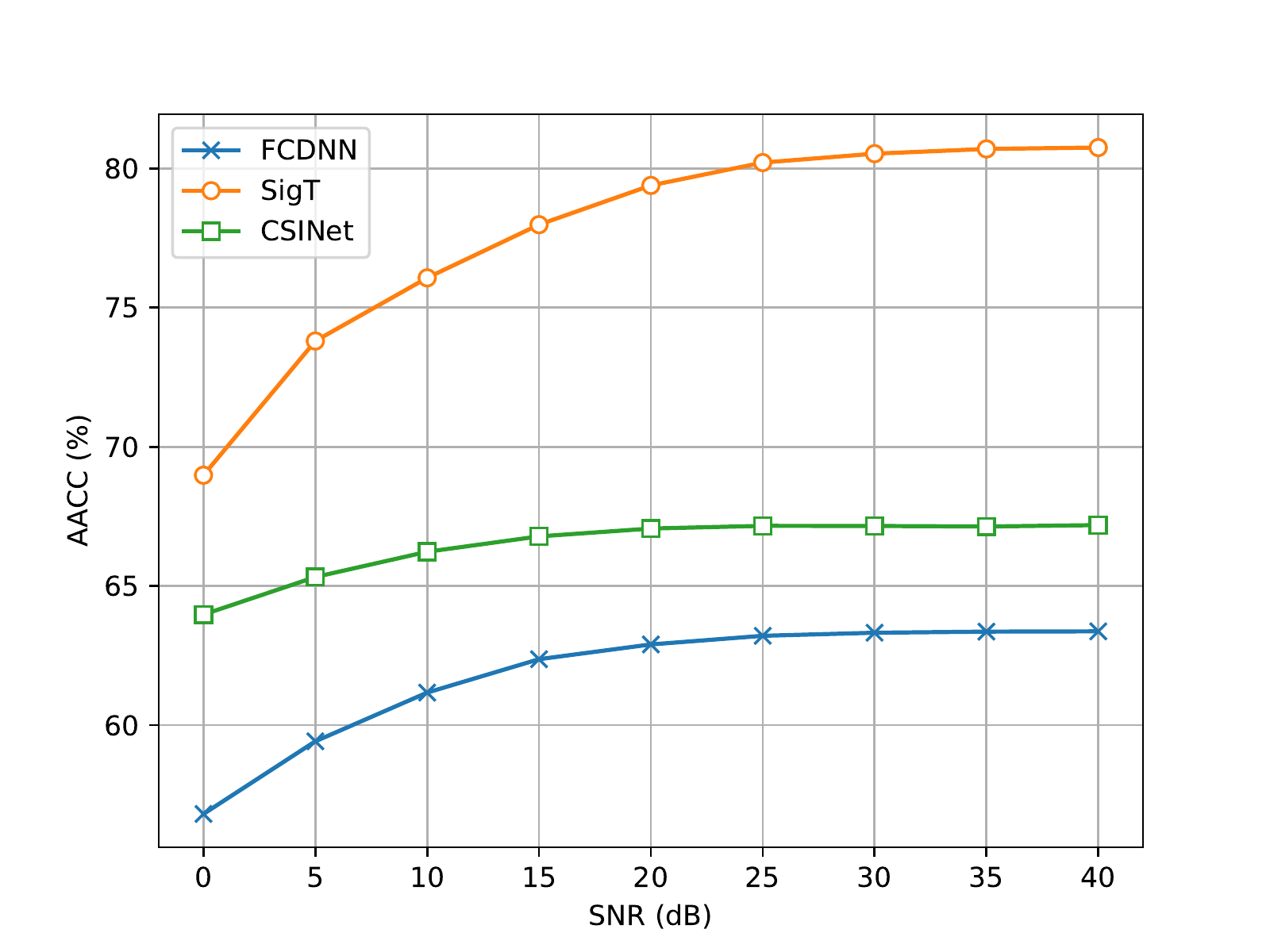}
  \vspace{-11pt}
  \caption {Test AACC for different signal noise ratio (SNR) for three DL-based models, i.e. FC-DNN, CSINet, and SigT. }
  \label{EXP_1}
  \vspace{-11pt}
\end{figure}
\begin{figure}[h]
  \centering
  \includegraphics[width=0.75\columnwidth]{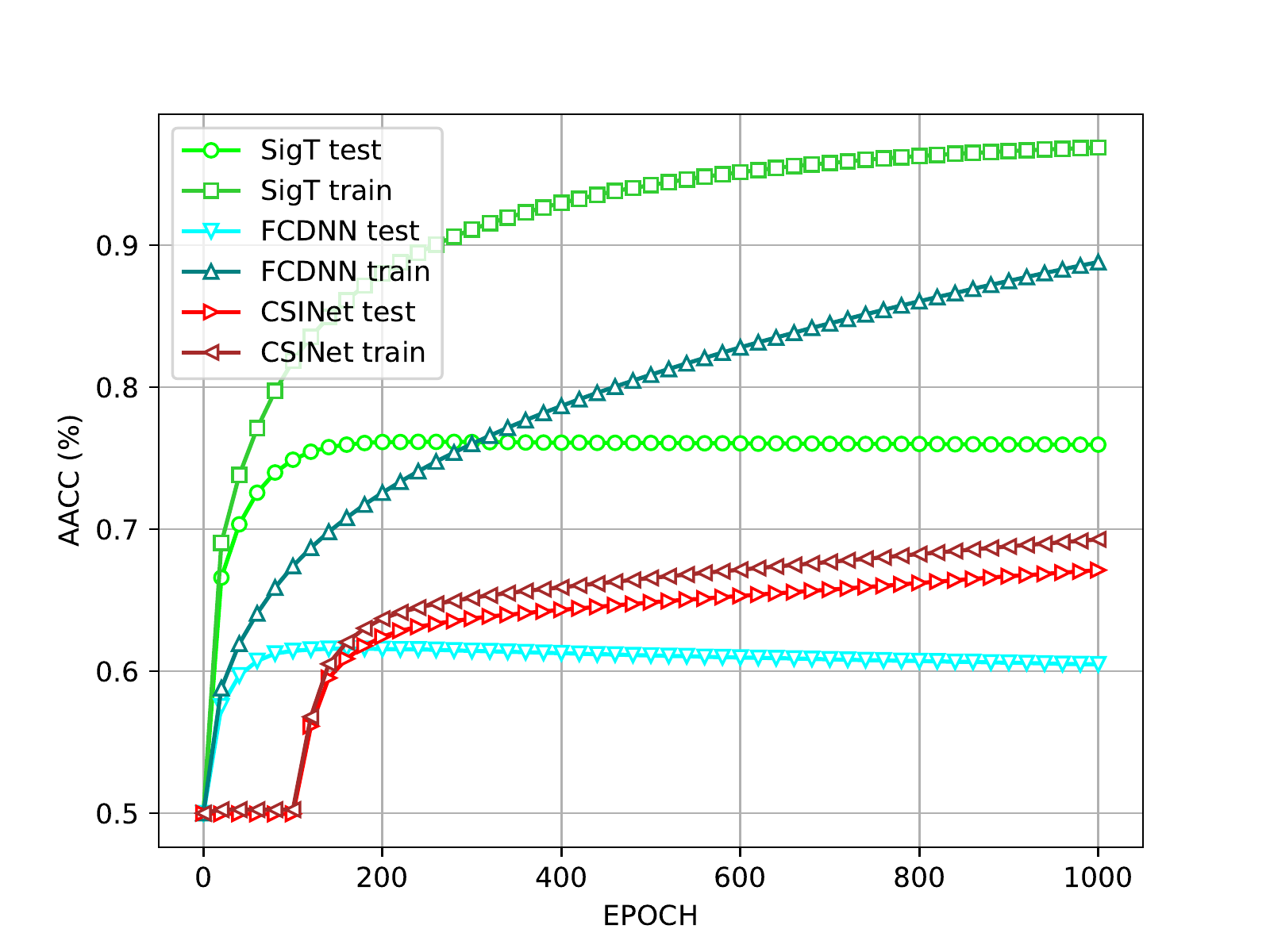}
  \vspace{-11pt}
  \caption {The AACC on the training set and test set for the three models.}
  \label{EXP_2}
  \vspace{-11pt}
\end{figure}

There are other hyperparameters that can affect the results of AACC. For simplicity, we set the optimizer to be adaptive momentum (Adam) with learning rate $0.0001$, $\beta_1 = 0.9,\quad \beta_2 = 0.999,\quad \epsilon=10^{-8}$. We do not apply the weight decay, i.e., regularization, unless otherwise specified. 

\subsection{Benchmark DL-Based Models}
The benchmark methods used in the experiment are the DL-based models in \cite{12,13}. We adapt the models to work with the MIMO-OFDM scenario and the used dataset. 

1) FC-DNN. This network consists $N_r$ independent MLPs, each following the design in \cite{12} but doubling the size of hidden layers.

2) CSINet. CSINet is stacked from RefineNet module, which is a residual CNN in \cite{13}. Flatten layer is added to match the signal to the ground truth.

In the experiment, we comprehensively evaluate the performance of our proposed model, SigT, and compare it with the benchmark DL-based models, namely MLP-structured FC-DNN and CNN-structured CSINet.

Fig. \ref{EXP_1} compares the AACC performance under different SNRs. During the training phase, we construct the training set that has the same SNR as the test set. 
The results show that SigT outperforms the other two models by a large margin since MSA can efficiently exploit the correlation among the receiving antennas. For SNR lower than 10 dB, all three models' AACC increases fast with the increase of SNR. This is because when the SNR becomes higher, the models can gradually learn the signal patterns which are covered by the noise when SNR is low. However, when SNR gets even higher, the error floor occurs mainly due to the generalization difficulty of the end-to-end model in the MIMO-OFDM setting.  

In Fig. \ref{EXP_2}, the performance on both the training set and test set is shown with 10 dB SNR. It can be seen that the proposed SigT achieves $14.9\%$ higher AACC than FC-DNN and $8.95\%$ higher AACC than CSINet. For FC-DNN, the large gap between training AACC and test AACC and the waning of test AACC after $400$ epochs reveals that the FC-DNN model faces the problem of overfitting.
Because the receiving system of MIMO-OFDM is too complicated and needs huge expert knowledge to fully understand. Thus, simply-designed DL networks cannot perform well on unseen signals. Similarly for the CSINet model, underfitting is normally derived from poor representative power. The proposed SigT model, in comparison, reaches a rather high AACC on the training and test set and grows faster than CSINet and FC-DNN. This high AACC and increasing rate indicate that SigT is able to capture the features and expert knowledge.



\begin{figure}[h]
  \centering
  \includegraphics[width=0.75\columnwidth]{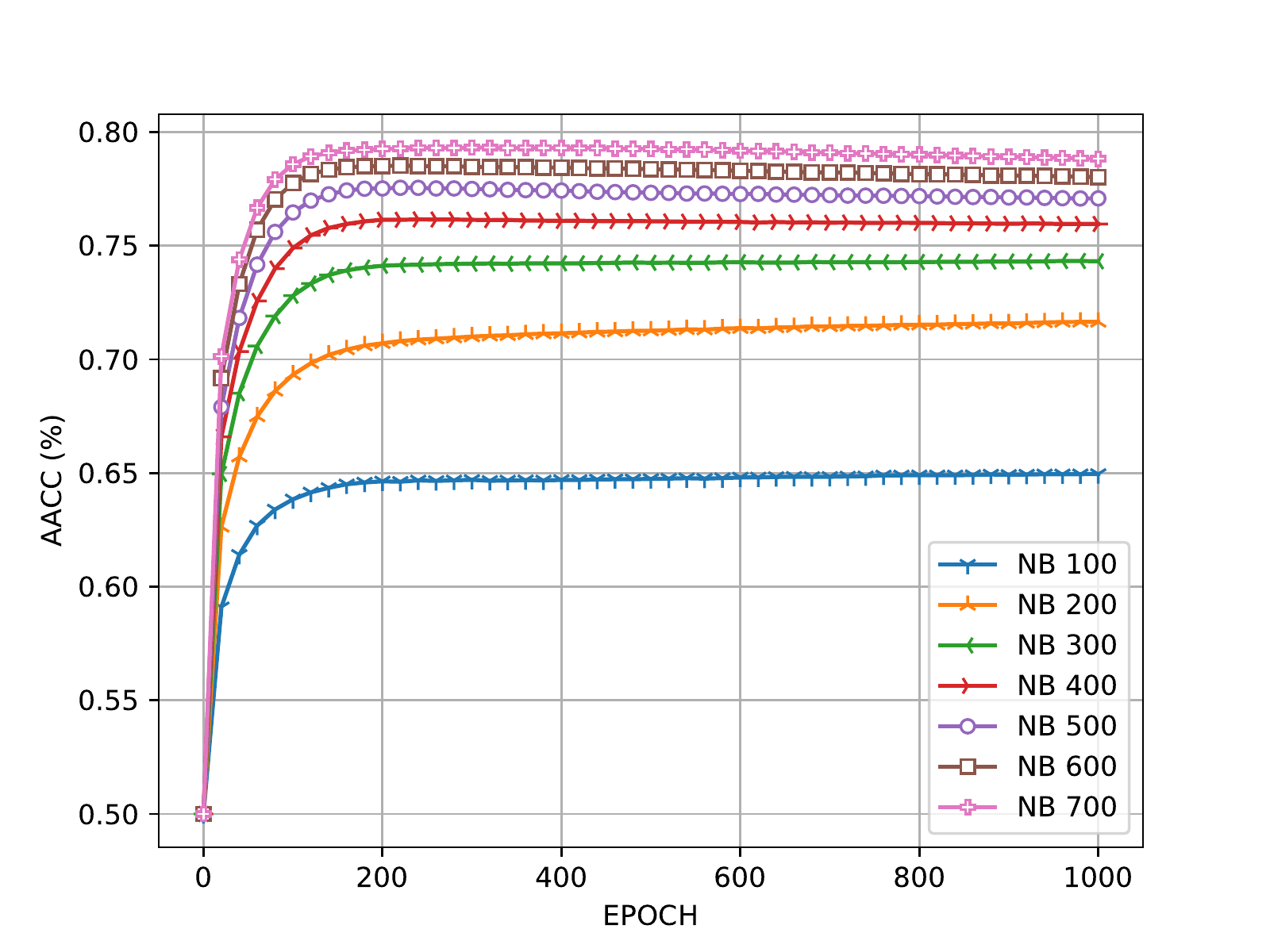}
  \vspace{-11pt}
  \caption {Test AACC of SigT for different $NB$ with $SNR=10$ dB}
  \label{EXP_3}
  \vspace{-11pt}
\end{figure}

The relation between the training set size and the model performance is shown in Fig. \ref{EXP_3}. In this simulation, we construct a batch with batch size equals $64$ training sample signals, and test the performance of SigT on different number of training signals by changing the number of batches, i.e. $num\_signals=\text{NB}\times 64$. As we can see from Fig. \ref{EXP_3}, the AACC on test set grows rapidly with the increase of NB when NB is small and starts to saturate when NB gets larger. This shows that simply increasing the size of the training set does not always result in improved learning performance.

\subsection{Ablation Study}
The ablation study of the proposed SigT model is conducted in this part.

The difference between the pooling layer and the convolutional layer is that the convolutional layer serves as a linear combination of features and is able to capture more local information and the pooling layer only serves to reduce the dimension of features. As a result, it is not unexpected that switching the pooling layer to the convolutional layer improves the performance of the model. As can be seen in the TABLE \ref{tab1e}, SigT with convolutional layers reaches higher AACC on both training set and test set than that with pooling layers.

\begin{figure}[h]
  \centering
  \includegraphics[width=0.75\columnwidth]{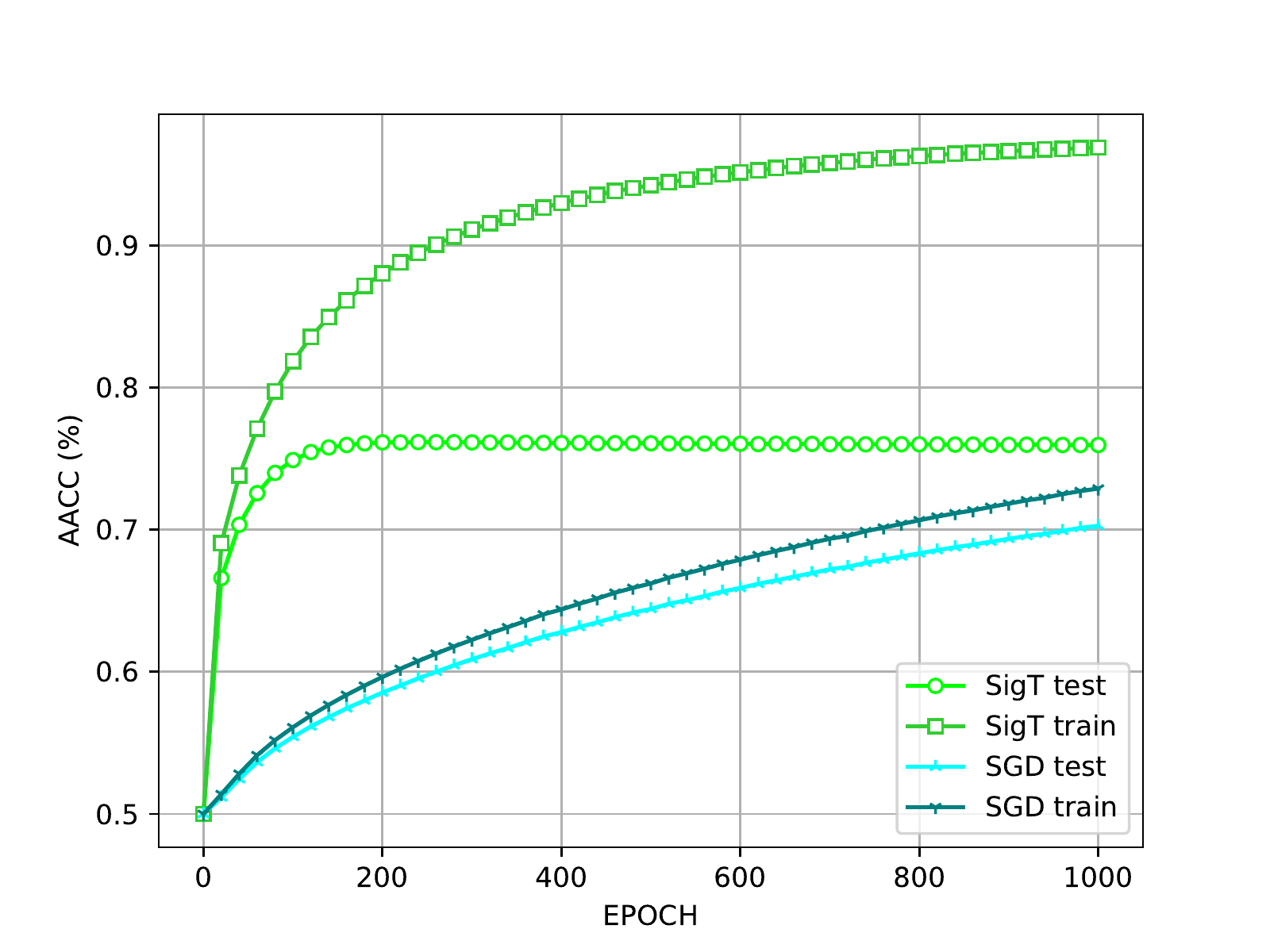}
  \vspace{-11pt}
  \caption {The AACC on the training set and test set for three models.}
  \label{EXP_4}
  \vspace{-10pt}
\end{figure}


Dropout is normally used for preventing overfitting by randomly masking part of neural units in MLP. However, such mask strategy would reduce the representative power when the size of hidden layers is not overly large. As shown in TABLE \ref{tab1e}, we apply dropout on both conv-based SigT and pool-based SigT's MLP head, the performance is lower than not applying dropout. Such results demonstrate that SigT performs well enough and can learn knowledge from training instead of just fitting a mapping function.

Different optimizers are also compared, where we explore two commonly used optimization functions, namely stochastic gradient descent (SGD) and adaptive momentum (Adam). As shown in Fig. \ref{EXP_4}, the model using Adam converges more quickly and reaches higher AACC. In comparison, the model using SGD not only converges more slowly but also performs poorly.





At last, to directly show the representative power of the transformer model, we test the performance of another widely used RNN model, long short-term memory (LSTM), as a backbone and retain the remaining part the same as SigT. To get the best results of LSTM, we apply convolutional layers, apply no dropout method, and use Adam to train $1000$ epochs. We can see that the LSTM-backbone model can hardly learn anything. With training AACC and test AACC far lower than SigT, we conclude that our model is more representative than other models.

\section{Conclusion}
In this paper, we have designed an end-to-end DL-based MIMO-OFDM receiver, named SigT, to efficiently detect and demodulate the receiving signal. The proposed SigT utilizes the transformer model to learn the knowledge of correlations among the receiving antennas to relieve the zero-shot problem confronted in end-to-end learning models. The experiment results have shown that the proposed SigT achieves the highest performance compared with the other two benchmark end-to-end DL-based models in terms of signal recovery accuracy. 
The proposed SigT can serve as a performance baseline for future study of learning-based MIMO-OFDM receiver design. In future work, we will study the dynamic mechanisms in DL models to adapt the DL-based receiver to dynamic environments.

\section*{Acknowledgement}
This work was supported by the National Key Research and Development Program of China (2020YFB1807700), the National Natural Science Foundation of China (NSFC) under Grant No. 62071356, and the Fundamental Research Funds for the Central Universities under Grant No. JB210113.

\ifCLASSOPTIONcaptionsoff
  \newpage
\fi

\bibliography{ref}

\begin{thebibliography}{10}
\providecommand{\url}[1]{#1}
\csname url@samestyle\endcsname
\providecommand{\newblock}{\relax}
\providecommand{\bibinfo}[2]{#2}
\providecommand{\BIBentrySTDinterwordspacing}{\spaceskip=0pt\relax}
\providecommand{\BIBentryALTinterwordstretchfactor}{4}
\providecommand{\BIBentryALTinterwordspacing}{\spaceskip=\fontdimen2\font plus
\BIBentryALTinterwordstretchfactor\fontdimen3\font minus
  \fontdimen4\font\relax}
\providecommand{\BIBforeignlanguage}[2]{{%
\expandafter\ifx\csname l@#1\endcsname\relax
\typeout{** WARNING: IEEEtran.bst: No hyphenation pattern has been}%
\typeout{** loaded for the language `#1'. Using the pattern for}%
\typeout{** the default language instead.}%
\else
\language=\csname l@#1\endcsname
\fi
#2}}
\providecommand{\BIBdecl}{\relax}
\BIBdecl

\bibitem{larsson2014massive}
E.~G. Larsson, O.~Edfors, F.~Tufvesson, and T.~L. Marzetta, ``Massive mimo for
  next generation wireless systems,'' \emph{IEEE communications magazine},
  vol.~52, no.~2, pp. 186--195, 2014.

\bibitem{stuber2004broadband}
G.~L. Stuber, J.~R. Barry, S.~W. Mclaughlin, Y.~Li, M.~A. Ingram, and T.~G.
  Pratt, ``Broadband mimo-ofdm wireless communications,'' \emph{Proceedings of
  the IEEE}, vol.~92, no.~2, pp. 271--294, 2004.

\bibitem{NIPS2012_c399862d}
A.~Krizhevsky, I.~Sutskever, and G.~E. Hinton, ``Imagenet classification with
  deep convolutional neural networks,'' \emph{Advances in Neural Information
  Processing Systems}, vol.~25, pp. 1--9, 2012.

\bibitem{9882279}
L.~Ma, N.~Cheng, X.~Wang, R.~Sun, and N.~Lu, ``On-demand resource management
  for 6g wireless networks using knowledge-assisted dynamic neural networks,''
  in \emph{ICC 2022 - IEEE International Conference on Communications}, 2022,
  pp. 1--6.

\bibitem{wang2022joint}
X.~Wang, L.~Fu, N.~Cheng, R.~Sun, T.~Luan, W.~Quan, and K.~Aldubaikhy, ``Joint
  flying relay location and routing optimization for 6g uav--iot networks: A
  graph neural network-based approach,'' \emph{Remote Sensing}, vol.~14,
  no.~17, p. 4377, 2022.

\bibitem{10}
H.~He, C.-K. Wen, S.~Jin, and G.~Y. Li, ``Model-driven deep learning for mimo
  detection,'' \emph{IEEE Transactions on Signal Processing}, vol.~68, pp.
  1702--1715, 2020.

\bibitem{11}
H.~Ye, G.~Y. Li, and B.-H. Juang, ``Power of deep learning for channel
  estimation and signal detection in ofdm systems,'' \emph{IEEE Wireless
  Communications Letters}, vol.~7, no.~1, pp. 114--117, 2018.

\bibitem{13}
J.~Guo, X.~Li, M.~Chen, P.~Jiang, T.~Yang, W.~Duan, H.~Wang, S.~Jin, and Q.~Yu,
  ``Ai enabled wireless communications with real channel measurements: Channel
  feedback,'' \emph{Journal of Communications and Information Networks},
  vol.~5, no.~3, pp. 310--317, 2020.

\bibitem{12}
P.~Jiang, T.~Wang, B.~Han, X.~Gao, J.~Zhang, C.-K. Wen, S.~Jin, and G.~Y. Li,
  ``Ai-aided online adaptive ofdm receiver: Design and experimental results,''
  \emph{IEEE Transactions on Wireless Communications}, vol.~20, no.~11, pp.
  7655--7668, 2021.

\bibitem{8715649}
J.~Liu, K.~Mei, X.~Zhang, D.~Ma, and J.~Wei, ``Online extreme learning
  machine-based channel estimation and equalization for ofdm systems,''
  \emph{IEEE Communications Letters}, vol.~23, no.~7, pp. 1276--1279, 2019.

\bibitem{8214233}
S.~Dörner, S.~Cammerer, J.~Hoydis, and S.~t. Brink, ``Deep learning based
  communication over the air,'' \emph{IEEE Journal of Selected Topics in Signal
  Processing}, vol.~12, no.~1, pp. 132--143, 2018.

\bibitem{gao2018comnet}
X.~Gao, S.~Jin, C.-K. Wen, and G.~Y. Li, ``Comnet: Combination of deep learning
  and expert knowledge in ofdm receivers,'' \emph{IEEE Communications Letters},
  vol.~22, no.~12, pp. 2627--2630, 2018.

\bibitem{NIPS2013_2d6cc4b2}
R.~Socher, M.~Ganjoo, C.~D. Manning, and A.~Ng, ``Zero-shot learning through
  cross-modal transfer,'' vol.~26, 2013.

\bibitem{vaswani2017attention}
A.~Vaswani, N.~Shazeer, N.~Parmar, J.~Uszkoreit, L.~Jones, A.~N. Gomez,
  {\L}.~Kaiser, and I.~Polosukhin, ``Attention is all you need,''
  \emph{Advances in neural information processing systems}, vol.~30, 2017.

\bibitem{dosovitskiy2020image}
A.~Dosovitskiy, L.~Beyer, A.~Kolesnikov, D.~Weissenborn, X.~Zhai,
  T.~Unterthiner, M.~Dehghani, M.~Minderer, G.~Heigold, S.~Gelly \emph{et~al.},
  ``An image is worth 16x16 words: Transformers for image recognition at
  scale,'' \emph{arXiv preprint arXiv:2010.11929}, 2020.

\end{thebibliography}
\bibliographystyle{IEEEtran}

\end{document}